\documentclass[11pt]{article}
\usepackage{amsfonts}
\usepackage{amssymb,epic,eepic}
\usepackage{graphics}
\usepackage[dvips]{graphicx}
\usepackage[dvips]{hyperref}
\usepackage{epsfig}

\def\ThmNum{yes}

\newfont{\bbf}{msbm10 at 12pt}
\newfont{\bbfsm}{msbm10 at 9pt}

\def\idn{{\textbf{1}}}

\def\N{\mathbb {N}}

\def\Z{\mathbb {Z}}
\def\R{\mathbb {R}}

\def\phi1{\phi}
\def\phi{\varphi}

\def\theta{\vartheta}

\ifx\ThmNum\undefined
      \newtheorem{theorem}{Theorem}[section]
      
\else
      \newtheorem{theorem}{Theorem}
      
\fi

\def\proof{\par\medskip\noindent {\sc Proof. }}

\def\proofof #1 {\par\medskip\noindent {\sc Proof of #1. }}

\def\sketchof #1 {\par\medskip\noindent {\sc Sketch of proof of #1. }}
\def\Box{\framebox[10pt]{\rule{0pt}{3pt}}}
\def\nix{\rule{0pt}{2pt}}
\def\qed{\qedd\par\medskip\noindent}
\def\qedd{\nix\nolinebreak\hfill\hfill\nolinebreak$\Box$}

\def\reminder #1 {{\sf #1}}
\def\hide #1 {}

\newcommand{\hr}{\mathcal{H}}
\newcommand{\I}{\mathcal{I}^{(n)}}
\newcommand{\bb}{\mathbf{b}}

\title{Exponential error rates in multiple state discrimination on a  quantum spin chain}

\begin{document}

\author{ Michael Nussbaum$^{1}$, Arleta Szko\l a$^{2}$\\ \\
 $^{1}${\footnotesize Department of Mathematics, Cornell University }
 \\ {\footnotesize Ithaca NY, 14853, USA } \\ {\footnotesize e-mail:
 nussbaum@math.cornell.edu} \\ \\ $^{2}${\footnotesize Max Planck
 Institute for Mathematics in the Sciences}\\ {\footnotesize
 Inselstrasse 22, 04103 Leipzig, Germany} \\ {\footnotesize e-mail:
 szkola@mis.mpg.de} }

\maketitle



\begin{abstract}
We consider decision problems on finite sets of hypotheses represented
by pairwise different shift-invariant states on a quantum spin
chain. The decision in favor of one of the hypotheses is based on
outputs of generalized measurements performed on local states on
blocks of finite size. We assume existence of the mean quantum
Chernoff distances of any pair of states from the given set and refer
to the minimum of them as the mean generalized quantum Chernoff
distance.

We establish that this minimum specifies an asymptotic bound on the
exponential decay of the averaged probability of rejecting the true
state in increasing block size, if the mean quantum Chernoff distance
of any pair of the hypothetic states is achievable as an asymptotic
error exponent in the corresponding binary problem. This assumption is
in particular fulfiled by shift-invariant product states
(i.i.d. states).

Further, we provide a constructive proof for the existence of a
sequence of quantum tests in increasing block size, which achieves an
asymptotic error exponent which is equal to the mean generalized
quantum Chernoff distance of the given set of states up to a factor,
which depends on the set itself. It can be arbitrary close to $1$ and
is not less than $1/m$ for $m$ being the number of different pairs of
states from the set considered.

\end{abstract}
\section{Introduction}\label{sec:Intro}
In the series of papers \cite{NSz}, \cite{Audenaert}, \cite{ANSV} the
decision problem between two density operators associated to quantum
states of a finite quantum system has been solved in the setting of
asymptotic quantum hypothesis testing -for some earlier useful results
obtained in this context see also~\cite{Kargin} and
~\cite{hayashi}. There decisions in favor of one of the two hypothetic
states appearing with an a priori probability strictly larger than
zero are based on outcomes of generalized measurements performed on a
finite number of copies of the quantum system, where the corresponding
quantum state is associated to a tensor product of one of the two
hypothetic density operators. The limit of a large number of copies
corresponds to a shift-invariant product state on a quantum spin
chain. According to~\cite{NSz},~\cite{Audenaert},~\cite{ANSV} it turns
out that there is a quantum version of the Chernoff distance defined
for pairs of hypothetic density operators, which specifies the best
asymptotic exponential decay of the averaged probability of rejecting
the true quantum state.  This is in analogy to results from classical
asymptotic hypothesis testing.

A canonical extension of the binary decision problem refers to a
finite number of hypotheses. In the setting of classical asymptotic
multiple hypothesis testing, where the hypotheses are represented by
probability distributions, the best asymptotic error exponent is equal
to the generalized Chernoff distance, see~\cite{Salikhov}. In our
recent work \cite{NSz2}, in analogy to the classical definition given
in~\cite{Salikhov}, we have introduced the \textit{generalized quantum
Chernoff distance} of a finite set of density operators as the minimum
of the binary quantum Chernoff distances over all possible pairs of
different hypothetic density operators. We could identify this minimum
as a bound on asymptotic error exponents in corresponding multiple
quantum hypothesis testing and establish that it is achievable in the
special case where the hypotheses are represented by pure quantum
states. For completeness we want to mention that there is a wide
literature treating the related problem of optimal multiple state
discrimination in a finite, i.e. non asymptotic setting,
cf.~\cite{Yuen},~\cite{buch-Holevo},~\cite{Holevo},~\cite{KRS},~\cite{Tyson},~\cite{Barett},~\cite{Gen}. The
optimal discrimination between exactly two density operators has been
completely solved by Helstrom and Holevo,
see~\cite{buch-Helstrom},~\cite{Holevo2}.

In the presence of correlations among the single quantum systems of a
spin chain the hypothetic states are represented by (in an appropriate
sense) compatible sequences, in an increasing block size, of density
operators in respective local algebras of observables. Several special
cases of hypotheses represented by correlated quantum states on spin
chains has been investigated by Hiai et. al in a series of
papers~\cite{hiai-mosonyi-ogawa},~\cite{hiai-mosonyi-ogawa2},~\cite{hiai-mosonyi-ogawa-fannes}. There
the quantum Chernoff distance of two density operators has been
replaced by the mean quantum Chernoff distance of two shift-invariant
states on a spin chain, which, roughly speaking, is defined as the
asymptotic rate of quantum Chernoff distances of the pairs of local
quantum states, if the corresponding limits exist. This is in line
with other well-established extensions of entropic quantities to the
case of shift-invariant correlated states on a spin chain; compare the
concepts of mean quantum relative entropy~\cite{hiai-petz} and mean
quantum entropy/quantum entropy rate~\cite{invent-paper}. From our
point of view the most relevant result among
~\cite{hiai-mosonyi-ogawa},~\cite{hiai-mosonyi-ogawa2},~\cite{hiai-mosonyi-ogawa-fannes}
is given in~\cite{hiai-mosonyi-ogawa}. It identifies a class of
shift-invariant states on a quantum spin chain, which is characterized
by a factorization property, as a domain where the mean binary
Chernoff distances exist and specify the best asymptotically
achievable error exponents in corresponding binary decision
problems. Note that similar classes of correlated states with
appropriate factorization property have been shown to permit
(classical and quantum) Sanov type theorems, which resolve some
related asymmetric decision problems, cf.~\cite{corr-q-sanov}.

In this paper we define the \textit{mean generalized quantum Chernoff
distance} of finite sets of pairwise different shift-invariant quantum
states on a spin chain as the minimum of the mean quantum Chernoff
distances of all the possible quantum state pairs. Notice that the
minimum is well-defined on the set of shift-invariant quantum states,
where all the binary quantum Chernoff distances exist, i.e. in
particular on both the set of shift-invariant product states and the
strictly larger set of shift-invariant states fulfilling the
factorization assumption as specified in \cite{hiai-mosonyi-ogawa}. We
point out that in the case of shift-invariant product states the mean
generalized quantum Chernoff distance coincides with the generalized
quantum Chernoff distance of the corresponding density operators
associated to the local states on the blocks of size $1$.

Extending the result presented in Theorem 1 of our previous
paper~\cite{NSz2}, we show that the mean generalized quantum Chernoff
distance, if it exists for a given finite set of shift-invariant
states, specifies a bound on the exponential decay of the averaged
error probability in corresponding multiple state
discrimination. Here, again, we assume that each of the hypothetic
states appears with an a priori probability strictly larger than
zero. As our main contribution we establish that an exponential decay,
i.e. a strictly positive asymptotic error exponent, is indeed
achievable in multiple state discrimination. To the best of our
knowledge this has not been shown so far apart from the case of
\textit{two} hypotheses, cf.~\cite{ANSV},~\cite{hiai-mosonyi-ogawa},
and the special case of multiple pure (i.i.d.) state discrimination,
cf.~\cite{NSz2}. More precisely, we construct a sequence of quantum
tests for the set of hypothetic local states, such that the
exponential decay of the averaged error probability in increasing
block size is equal to the mean generalized quantum Chernoff distance
up to a factor, which depends on the configuration of the states. The
factor can be arbitrary close to $1$. In the worst case, where all the
involved binary mean Chernoff distances are equal, it is equal to
$1/{r \choose 2}$, where $r$ is the number of different hypothetic
states. Our construction represents an appropriate blockwise
combination of the optimal quantum tests of the associated asymptotic
binary decision problems.

The outline of our paper is as follows:
\begin{itemize}
	\item In Section \ref{sec:Notations} we introduce our
	notations, explain shortly the mathematical framework of a
	quantum spin chain and its state space, present the
	definitions of the here relevant Chernoff type distances and
	finally we are in the position to state precisely our main
	results in Theorems~\ref{thm:gen-Chernoff-lower-bound}
	and~\ref{thm:exp-decay}.

	\item The proof of Theorem~\ref{thm:gen-Chernoff-lower-bound},
	which adopts the idea of the proof of Theorem 1 from our
	previous paper~\cite{NSz2}, is given in
	Section~\ref{sec:lower-bound}.

	\item Section~\ref{sec:exp-decay} contains a construction of
	quantum tests for multiple states on a quantum spin chain,
	which -subject to the assumptions of
	Theorem~\ref{thm:exp-decay}- achieves an asymptotic error
	exponent equal to the mean generalized quantum Chernoff
	distance up to a factor depending on the set of states
	itself. This proves our main Theorem~\ref{thm:exp-decay}.
\end{itemize}

\section{Notations and main results}\label{sec:Notations}
Let $\hr$ be a complex Hilbert space with $\dim \hr =d < \infty$ and
$\mathcal{A} $ be a unital $C^*$-subalgebra of linear operators on
$\hr$. For each finite subset $\Lambda \subset \Z$ denote by
$\mathcal{A}_{\Lambda}$ the tensor product $\bigotimes_{i \in \Lambda}
\mathcal{A}$, which is a $C^*$-subalgebra of linear operators on
$\bigotimes_{i \in \Lambda} \hr$. The construction of quasi-local
$C^*$-algebras $\mathcal{A}^{\infty}$ formalizes the limit of
$\mathcal{A}_{\Lambda}$, as $\Lambda$ tends to be $\Z$, compare
\cite{buch-Ruelle} or \cite{buch-Bratteli}. 

The state space $\mathcal{S}(\mathcal{A}^{\infty})$ of
$\mathcal{A}^{\infty}$ consists of positive linear functionals $\omega:
\mathcal{A}^{\infty} \to \mathbb{C}$ fulfilling the normalization
condition $\omega (\idn)=1$, where $\idn$ denotes the identity in
$\mathcal{A}^{\infty}$. Each $ \omega
\in\mathcal{S}(\mathcal{A}^{\infty})$ corresponds one-to-one to a
family of local states $\omega_{\Lambda}$, $\Lambda \subset \Z $ with
$|\Lambda|<\infty$, being restrictions of $\omega $ onto
$\mathcal{A}_{\Lambda}$, respectively. We are primarily interested in
the convex subset $\mathcal{T}(\mathcal{A}^{\infty})$ of
shift-invariant states on $\mathcal{A}^{\infty}$. Note that the
shift-invariance implies that for any $\Lambda_1, \Lambda_2 \subset \Z
$ of equal size, i.e. with $|\Lambda_1|=|\Lambda_2|$, we can identify
the corresponding restrictions $\omega_{\Lambda_1}$ and
$\omega_{\Lambda_2}$ of $\omega \in
\mathcal{T}(\mathcal{A}^{\infty})$. It follows that a shift-invariant
state $\omega$ is determined by a sequence of local states
$\omega^{(n)}$, $n \in \N$, on
$\mathcal{A}^{(n)}:=\mathcal{A}_{[1,n]}$, respectively. For each $n
\in \N$ the associated density operator $\rho^{(n)} \in
\mathcal{A}^{(n)}$ satisfies $\omega^{(n)} (a)= \textrm{tr }
\rho^{(n)} a$ for all $a\in \mathcal{A}^{(n)}$.

Let $\Sigma$ be a finite set of states $\omega_i \in
\mathcal{T}(\mathcal{A}^{\infty})$, $i=1, \dots, r$, representing the
hypotheses $H_i$, respectively. We can identify $\Sigma $ with the
sequence $\Sigma^{(n)}$, $n \in \N$, of sets of associated density
operators $\rho^{(n)}_i$, $i=1,\dots, r$, in $\mathcal{A}^{(n)}$,
respectively. For each $n \in\N$ let $E^{(n)}=\{E_i^{(n)}\}_{i=1}^r$
be a positive operator valued measure (POVM) in $\mathcal{A}^{( n)}$,
i.e. each $E_i^{(n)}$, $i=1,\dots, r$, is a self-adjoint element of
$\mathcal{A}^{(n)}$ with $E_i^{(n)} \geq 0$ and $ \sum_{i=1}^r
E_i^{(n)} = \idn$. The POVMs $E^{(n)}$ determine generalized
measurements. By identifying the measurement outcome corresponding to
$E^{(n)}_i$, $i=1,\dots,r$, with the hypothesis $H_i \sim
\rho^{(n)}_i$, respectively, they describe quantum tests for
discrimination between the quantum states associated to density
operators from $\Sigma^{(n)}$, or simply {\em quantum tests for
$\Sigma^{(n)}$}. If $\omega_i$ happens to be the true state then the
corresponding {\em individual success probability} is given by
\begin{eqnarray}
	\textrm{Succ}_i(E^{(n)}):=\textrm{tr } [\rho_i^{(n)}
	E_i^{(n)}].
\end{eqnarray}
and consequently the {\em individual error probability} is
\begin{eqnarray}
	\textrm{Err}_i(E^{(n)}):=\textrm{tr } [\rho_i^{(n)} (\idn -
	E_i^{(n)})].
\end{eqnarray}
It refers to the situation when $H_i$ is rejected. Assuming $0 < p_i
< 1$, $i=1,\dots, r$, with $\sum_{i=1}^r p_i=1$ to be the prior
distribution on the given set of $r$ hypotheses the {\em averaged
error probability} is given by
\begin{eqnarray}
	\textrm{Err}(E^{(n)})= \sum_{i=1}^r p_i \textrm{tr }
	[\rho_i^{(n)} (\idn - E_i^{(n)})].
\end{eqnarray}   
If the limit $\lim_{n \to \infty} -\frac{1}{n}\log
\textrm{Err}(E^{(n)})$ exists, we refer to it as the {\em asymptotic error
exponent}. Otherwise we have to consider the corresponding $\limsup$
and $\liminf$ expressions.
\\
\\
For two density operators $\rho_1$ and $\rho_2$ the \textit{quantum
Chernoff distance} is defined by
\begin{eqnarray}\label{def:q-Chernoff-dist}
	\xi_{QCB}(\rho_1, \rho_2):=-\log\inf_{0\leq s \leq 1}
	\textrm{tr } \rho_1^{1-s} \rho_2^{s}.
\end{eqnarray}
It specifies the optimal achievable asymptotic error exponent in
discriminating between $\rho_1$ and $\rho_2$, compare \cite{NSz},
\cite{Audenaert}, \cite{ANSV}. 
Quantum tests with minimal averaged error probability for a pair of
density operators $\rho_1$ and $ \rho_2$ on the same Hilbert space
$\hr$ are well-known to be given by the respective {\em
Holevo-Helstrom projectors}
\begin{eqnarray}
	\Pi_{1}&:=& \textrm{supp }(\rho_1 -\rho_2)_+,\\ \Pi_{2}&:=&
	\textrm{supp }(\rho_2 -\rho_1)_+ = \idn- \Pi_1,
\end{eqnarray}
where $\textrm{supp }a$ denotes the support projector of a
self-adjoint operator $a$, while $a_+$ means its positive part,
i.e. $a_+= (|a|+a)/2$ for $|a|:= (a^*a)^{1/2}$, see \cite{Holevo2},
\cite{buch-Helstrom}. The Holevo-Helstrom projectors generalize the
maximum likelihood tests for two probability distribution. This can be
verified by letting $\rho_1$ and $\rho_2$ be two commuting
density matrices. 

For a set $\Sigma=\{\rho_i\}_{i=1}^r$ of density operators in
$\mathcal{A}$, where $r >2$, we have introduced in \cite{NSz} the {\em
generalized quantum Chernoff distance}
\begin{eqnarray}\label{def:q-multiple-Chernoff-dist}
	\xi_{QCB}(\Sigma):=\min \{\xi_{QCB}(\rho_i, \rho_j):\ 1\leq
	i<j\leq r\}.
\end{eqnarray}  
This is in full analogy to the classical case where the hypotheses are
represented by probability distributiosn $P_i$, $i=1,\dots, r$, on a
finite sample space $\Omega$, see~\cite{Salikhov}.

In \cite{hiai-mosonyi-ogawa} the \textit{mean quantum Chernoff
distance} between two states $\omega_1$ and $\omega_2$ in
$\mathcal{T}(\mathcal{A}^{\infty})$, each of them  
corresponding one-to-one to the respective sequences $\{\rho_i^{(n)}\}_{n \in
\N}$, $i=1,2$, of density operators in corresponding local algebras $\mathcal{A}^{(n)}$, has been defined by
\begin{eqnarray}\label{def:mean-q-Chernoff-dist}
	\bar \xi_{QCB}(\omega_1, \omega_2 ):=\sup_{0\leq s\leq 1}\bar
	\xi_{QCB}^{(s)}(\omega_1, \omega_2 )
\end{eqnarray}
if the limits
\begin{eqnarray}
	\bar \xi_{QCB}^{(s)}(\omega_1, \omega_2 ):=\lim_{n \to \infty}
	-\frac{1}{n} \log \textrm{tr }[(\rho_1^{(n)})^{1-s}
	(\rho_2^{(n)})^{s}],
\end{eqnarray}
exist for $0\leq s\leq 1$. Note that in the special case where
$\omega_1$ and $\omega_2$ both are shift-invariant product states,
i.e. $\rho_i^{(n)}=\rho_i^{\otimes n}$ for all $n \in \N$, we have the
relation $\bar \xi_{QCB}(\omega_1, \omega_2 )=
\xi_{QCB}(\rho_1,\rho_2)$, i.e. the mean quantum Chernoff distance
coincides with the quantum Chernoff distance of the associated density
operators $\rho_1$ and $\rho_2$ in $\mathcal{A}^{(1)}$.

Finally, for a set $\Sigma=\{\omega_i\}_{i=1}^r$ of states on
$\mathcal{A}^{\infty}$ where the mean quantum Chernoff distances $\bar
\xi_{QCB}(\omega_i,\omega_j)$ exist for all pairs
$(\omega_i,\omega_j)$ with $i\not= j$, we introduce the {\em mean
generalized quantum Chernoff distance}
\begin{eqnarray}\label{def:multiple-mean-q-Chernoff-dist}
	\bar \xi_{QCB}(\Sigma):= \min \{\bar \xi_{QCB}(\omega_i,
	\omega_j):\ 1\leq i<j\leq r\}.
\end{eqnarray}
In \cite{NSz2}, see Theorem 1 therein, we have shown that in the case
of multiple shift-invariant product states on $\mathcal{A}^{\infty}$
the generalized quantum Chernoff distance of the associated set of
local states on $\mathcal{A}^{(1)}$ provides a bound on asymptotically
achievable error exponent in the corresponding multiple quantum
hypothesis testing. Here we extend the statement to the case of
hypotheses being represented by elements from a class of
shift-invariant correlated quantum states on
$\mathcal{A}^{\infty}$. The bound is then given by the corresponding
mean generalized quantum Chernoff distance.
\begin{theorem}\label{thm:gen-Chernoff-lower-bound}
Let $r \in \N$ and $\Sigma=\{ \omega_i \}_{i=1}^r$ be a set of states
on $\mathcal{A}^{\infty}$ with respective prior probability $0 < p_i
< 1$. If for every $(i,j)$, $1\leq i < j\leq r$, the
mean quantum Chernoff distance $\xi_{QCB}(\omega_i,
\omega_j)$ exists and specifies the optimal asymptotic error
exponent in the corresponding binary quantum hypothesis testing, then
it holds for any sequence $E^{(n)}$, $n \in \N$ of POVMs for
$\Sigma^{(n)}$, respectively,
\begin{eqnarray}
	\limsup_{n \to \infty} -\frac{1}{n} \log \textrm{Err}(E^{(n)})
	\leq \bar \xi_{QCB}(\Sigma),
\end{eqnarray}
where $\bar \xi_{QCB}(\Sigma)$ denotes the mean generalized quantum
Chernoff distance defined by
(\ref{def:multiple-mean-q-Chernoff-dist}).
\end{theorem}
As already mentioned, the assumption of Theorem
\ref{thm:gen-Chernoff-lower-bound} above is in particular satisfied on
the set of shift-invariant product states on $\mathcal{A}^{\infty}$,
cf.~\cite{ANSV}. Moreover, it has been shown
in~\cite{hiai-mosonyi-ogawa}, that it is also fulfilled on a subset of
shift-invariant correlated states with certain lower and upper
factorization properties. More preciesely, for a corresponding
shift-invariant state $\omega \in \mathcal{T}(\mathcal{A}^{\infty})$
there exist constants $\alpha, \beta \in \R$, and an $m_0\in \N$ such
that for all $m\geq m_0$ and $k\in \N$ it holds
\begin{eqnarray*}
	\omega_{[1,km]}\geq \alpha^{k-1}\omega_{[1,m]}^{\otimes
	k},\qquad \omega_{[1,km]}\leq
	\beta^{k-1}\omega_{[1,m]}^{\otimes k},
\end{eqnarray*}       
where $\omega_{[1,m]}$ denotes the restriction of $\omega$ onto the
local subalgebra $\mathcal{A}_{[1,n]}\subset \mathcal{A}^{\infty}$
associated to the finite block $[1,n]$ of the lattice $\Z$, which
underlies the quantum spin chain. For more details on the
factorization property and nontrivial examples such as Gibbs states of
translation-invariant finite-range interactions and finitely
correlated states see~\cite{hiai-mosonyi-ogawa}.

According to Theorem 2 in \cite{NSz2}, in the special case of a finite
set of pure states on $\mathcal{A}^{(1)}$ the corresponding
generalized quantum Chernoff distance indeed is achievable as an
exponential decay of minimal averaged error probability in
discrimination between the associated shift-invariant product states
on $\mathcal{A}^{\infty}$. The following theorem states that in the
general case of arbitrary (i.e. possibly mixed) density operators in
$\mathcal{A}^{(1)}$ an exponential rate of decay is achievable. We
exhibit an exponent which equals to the generalized quantum Chernoff
distance up to a factor, where the factor depends on the set of states
considered. Moreover, a similar result holds in the case of
shift-invariant correlated states on $\mathcal{A}^{\infty}$ fulfilling
the assumptions of Theorem \ref{thm:gen-Chernoff-lower-bound}. Here we
find an exponent which equals the \textit{mean} generalized quantum
Chernoff distance up to a factor, where again the factor depends on
$\Sigma$.
\begin{theorem}\label{thm:exp-decay}
Let $\Sigma$ be a finite set consisting of hypotheses $\omega_i \in
\mathcal{T}(\mathcal{A}^{\infty})$, $i=1, \dots, r$, such that the
mean quantum Chernoff distances $\bar \xi_{QCB}(\omega_i, \omega_j)$,
$1\leq i<j\leq r$, exist, are greater than zero, and represent
achievable asymptotic error exponents in the corresponding binary
hypothesis testing problems. Then there exists a sequence of quantum
tests $\{ E^{(n)}\}_{n \in \N}$ for $\Sigma^{(n)}$, respectively, such
that the corresponding averaged error probabilities satisfy
\begin{eqnarray}\label{bound-thm}
	\liminf_{n \to \infty}- \frac{1}{n} \log \mbox{Err}(E^{(n)})
	\geq \bar \xi_{QCB}(\Sigma) \phi(\Sigma),
\end{eqnarray}
where 
\begin{eqnarray}\label{def:phi-factor}
	\phi(\Sigma):=\left(\sum_{1\leq j < i \leq r} \frac{\bar
	\xi_{QCB}(\Sigma)}{\bar
	\xi_{QCB}(\omega_i,\omega_j)}\right)^{-1}.
\end{eqnarray}
\end{theorem}
The factor $\phi(\Sigma)$ satisfies
\[
	\frac{1}{m}=\frac{1}{\sum_{1\leq j<i,1\leq i\leq
	r}1}\leq\phi( \Sigma)
	\leq\frac{1}{\frac{\bar\xi_{QCB}}{\bar\xi_{QCB}}}=1.
\]
As a result, we can claim that the mean quantum Chernoff bound
$\bar\xi_{QCB}(\Sigma)$ is attainable up to a factor
$\phi(\Sigma)$. This factor is close to $1$ if the pairwise mean
quantum Chernoff distance for the least favorable pair $(i^{*},j^{*})$
(i.e. $\bar\xi_{QCB}(\Sigma)=\bar\xi_{QCB}(\omega_{i^{*}},\omega_{j^{*}})$)
is sufficiently small compared to the pairwise mean quantum Chernoff
distance $\bar\xi_{QCB}(\omega_i,\omega_j)$ for all other pairs, in
other words, if the least favorable pair $(i^{*},j^{*})$ sufficiently
"stands out" with regard to its estimation difficulty. If the other
extreme holds, i.e. all $\bar\xi_{QCB}(\omega_i,\omega_j)$ are equal,
then $\phi(\Sigma)$ is equal to its lower bound $1/m$.
\section{A Chernoff type bound in multiple state discrimination}\label{sec:lower-bound}
In this section we show that the generalized mean quantum Chernoff
distance provides a bound on the asymptotically achievable error
exponent in multiple quantum hypotheses testing, where the hypotheses
are represented by states on $\mathcal{A}^{\infty}$, such that for any
pair of them the (binary) mean Chernoff distance exists, is greater
than zero, and specifies the asymptotically optimal error exponent in
the corresponding binary hypothesis testing problem.
\proof[Theorem \ref{thm:gen-Chernoff-lower-bound}]
Denote by $\textrm{Err}_i(E^{(n)})$ the individual error probability
pertaining to the case that the true hypothesis $H_i$ corresponding to
the $n$-block density operator $\rho_i^{(n)}\in \mathcal{A}^{(n)}$ is
rejected on the base of outcomes of the quantum test $E^{(n)}$ for
$\Sigma$. Fix any two indices $1\leq i<j \leq r$. For $n \in \N$ let
$A^{(n)}, B^{(n)}
\in \mathcal{A}^{(n)}$ be two positive operators such that
$A^{(n)}+B^{(n)}=\idn -E^{(n)}_i- E^{(n)}_j$. Then the positive
operators $\tilde{E_i}^{(n)}:= E_i^{(n)}+ A^{(n)}$ and
$\tilde{E_j}^{(n)}:= E^{(n)}_j + B^{(n)}$ represent a POVM
$\tilde{E}^{(n)}$ in $\mathcal{A}^{(n)}$, which we regard as a quantum
test for the pair $\{\rho^{(n)}_i, \rho^{(n)}_j\}$. We obtain for the
modified individual error probabilities
\begin{eqnarray*}
	\textrm{Err}_i(\tilde E^{(n)})= \textrm{tr } [\rho_i^{(n)}
	(\idn- \tilde{E_i}^{(n)})] \leq \textrm{tr } [\rho_i^{(n)}
	(\idn- E_i^{(n)})]=\textrm{Err}_i(E^{(n)}),
\end{eqnarray*}
and similarily $\textrm{Err}_j(\tilde E^{(n)}) \leq
\textrm{Err}_j(E^{(n)})$.  It follows a lower bound on the average
error probability with respect to the original tests $\{
E_i^{(n)}\}_{i=1}^r$:
\begin{eqnarray*}
	\textrm{Err}(E^{(n)}) = \frac{1}{r} \sum_{k=1}^r
	\textrm{Err}_k(E^{(n)}) \geq \frac{1}{r} \left(
	\textrm{Err}_i(E^{(n)}) + \textrm{Err}_j(E^{(n)})\right) \geq
	\frac{1}{r} \left( \textrm{Err}_i(\tilde E^{(n)}) +
	\textrm{Err}_j(\tilde E^{(n)}) \right),
\end{eqnarray*}
which implies
\begin{eqnarray*}
	\limsup_{n \to \infty} -\frac{1}{n} \log \textrm{Err}(E^{(n)})
	&\leq& \limsup_{n \to \infty} \frac{1}{n} \log r + \limsup_{n
	\to \infty} -\frac{1}{n} \log \left( \textrm{Err}_i(\tilde
	E^{(n)}) + \textrm{Err}_j(\tilde E^{(n)}) \right) \\ &=&
	\limsup_{n \to \infty} -\frac{1}{n} \log \frac{1}{2} \left(
	\textrm{Err}_i(\tilde{E}^{(n)}) +
	\textrm{Err}_j(\tilde{E}^{(n)}) \right)
\\ &\leq& \xi_{QCB}(\omega_i,
	\omega_j).
\end{eqnarray*}
Here the last inequality holds by assumption of the validity of the
quantum Chernoff theorem for binary hypothesis testing. Since the pair
of indices $(i,j)$ was choosen arbitrarily, the statement of the theorem
follows. \qed
\section{Exponential decay of the averaged error probability}\label{sec:exp-decay}
The main idea of the proof of Theorem \ref{thm:exp-decay} is a
blockwise application of the optimal quantum test for pairs of quantum
states from the given set $\Sigma$. More in detail, the construction
of our quantum test can be described as follows. Consider all pairs of
states $\omega_i,\omega_j$, $i \not= j$, and divide the $n$-block
density operator into blocks of unequal size. Each block will be used
for testing between a particular pair, and the size of the blocks is
chosen in such a way that pairs of states which are more difficult to
discriminate are assigned longer blocks (more sample size). Within
each block a quantum measurement is performed confirming to the pair
of states, creating a decision random variable with values in
$\{i,j\}$ (a ``vote'' for either $i$ or $j$). When the random
variables for all blocks are realized, a final decision is made in
favor of hypothesis $H_i \sim \omega_i$ if this hypothesis has the
most number of votes. This can be broken in any way, for instance by
considering the numerical rank of $i$.
\\ 
\\
It is easy to see that in the commuting case, where for each $n$-block
the corresponding hypothetic density operators commute, this method is
related to maximum likelihood, though it does not coincide. In the
commuting case, there is no need for blocking and a direct maximum
likelihood decision is better. In the quantum (noncommuting) case, the
Yuen-Kennedy-Lax (YKL) test is the appropriate generalization of
maximum likelihood, see~\cite{Yuen}. It has minimum error probability
for any $n$, and it is a conjecture that its risk asymptotics is
described by the generalized (multiple) quantum Chernoff distance. Our
construction by blocking yields a feasible quantum test which can be
near-optimal for certain configurations of states, in terms of the
(mean) generalized quantum Chernoff distance. In this cases, it
provides an upper risk bracket close to the Chernoff bound for the YKL
test.
\proof[Theorem \ref{thm:exp-decay}] 
Let $m:={r \choose 2}$. This equals the number of different pairs of
states in $\Sigma$. Since we are interested in the asymptotic
behaviour in $n$ there is no loss of generality assuming $n\geq m$.

The main idea is to divide the discrete interval $[1,n]=:\I$ into
disjoint subblocks $\I_k$, $k=1,\dots, m$, of length $n_k$ each, each
of them being associated to one of the $m$ different density operator
pairs $\{\rho_i^{( n_k)}, \rho_j^{(n_k)}\}$, $i \not= j$. In order to
make the correspondence between $\{\I_k\}_{k=1}^{m}$ and the set of
unordered pairs $\left\{\{\rho_i^{(n_k)}, \rho_j^{(n_k)}\}\right\}$
one-to-one, we define the mapping
\begin{eqnarray}\label{index-ordering}
	\left\{1,\dots, m\right\}\ni k \mapsto (k_1,k_2)
	\in \left\{1,\dots,r \right\}^2,
\end{eqnarray}  
which to each $k \in\{1,\dots, m\}$ assigns an ordered pair of
indices $(i,j)$ in their lexicographic order, for $1\leq i\leq r-1$
and $i< j \leq r$. Now, that the one-to-one mapping $k \leftrightarrow
\{i,j\}, i \not= j$ is specified, we write $n(i,j):=n_{k}$ for the
length of the subblock associated to the pair $\left\{ i,j\right\} $,
and for ease of notation we also set $n(i,j):=n(j,i)$ for $j<i$.  The
lengths $n_{k}$ which satisfy $\sum_{s=1}^{m}n_{k}=n$ will be left
unspecified for now; we will determine them later.

In this construction, each subbblock ${\mathcal{I}}_{k}^{(n)}$ is now
associated to a pair of density operators
\[
	{\mathcal{I}}_{k}^{(n)}\mapsto\left\{ \rho_{k_{1}}^{( n(i,j))}
	,\rho_{k_{2}}^{(n(i,j))}\right\} ,\qquad k=1,\dots,{m}.
\]
We intend to construct the quantum test $E^{(n)}$ for $\Sigma^{( n)}$
as a composition of quantum tests of minimal averaged error
probability for the different pairs $\{\rho_i^{(
n_k)},\rho_j^{(n_k)}\}$, $1\leq i < j \leq r$. Optimal binary quantum
tests for any block size $n$ are known to be given by the Holevo-Helstrom tests
$(P_{i,j}^{(n)}, P_{j,i}^{(n)})$, where
\begin{eqnarray*}
	P_{i,j}^{(n)}:=\textrm{supp }(\rho_i^{(n)}-
	\rho_j^{(n)})_+
\end{eqnarray*}
is the orthogonal projector associated to the density operator
$\rho_i^{(n)}$, while $P_{j,i}^{(n)}:=\idn-P_{i,j}^{(n)}$ is
associated to $\rho_j^{(n)}$. We will apply these for any pair
$\left\{ i,j\right\} $ in the corresponding block of given size
$n(i,j)$. More precisely, our construction of $E^{(n)}$ works as
follows. Let $\pi_0$ and $\pi_1$ be permutations given by
\begin{eqnarray*}
\pi_0(i,j):=(i,j)\quad\textrm{and}\quad \pi_1(i,j):=(j,i)
\end{eqnarray*}
for $(i,j) \in \N \times\N$, and define for any vector $\bb \in
\{0,1\}^{m}$ an $m$-fold tensor product
projector in $\mathcal{A}^{(n)}$
\begin{eqnarray}\label{def:P_bb}
	P_{\bb}^{(n)}:=\bigotimes_{k=1}^{m}
	P_{\pi_{b_k}(k_1,k_2)}^{(n_k)},
\end{eqnarray}
where $b_k\in\{0,1\}$ denotes the $k$th coordinate of $\bb$. Observe
that the orthogonal projectors $P_{\bb}^{(n)}$, $\bb \in \{0,1\}^{m}$,
define a decomposition of the identity $\idn_n$ in $\mathcal{A}^{(n)}$,
i.e.
\begin{eqnarray}
	\sum_{\bb \in \{0,1\}^{m}} P_{\bb}^{(n)}=\idn_n,
\end{eqnarray}
and in this sense they represent a POVM $\tilde E^{(n)}$ in
$\mathcal{A}^{(n)}$ with $2^{m}$ elements.
\\
\\
We want to modify $\tilde E^{(n)}$, such that it represents a POVM
consisting of fewer, namely $m$ positive elements. Subsequently, by associating each
of the newly defined $m$ elements to a different density operator from
$\Sigma^{(n)}$ we obtain a \textit{quantum test} for
$\Sigma^{(n)}$. For each $i \in \{ 1,\dots, r\}$ we introduce the
function
\begin{eqnarray}
	n_i: \{0,1\}^{m} &\rightarrow& \{0,\dots, r-1
	\},\nonumber \\ \bb &\mapsto& n_i(\bb):= |\{k:\
	\pi_{b_k}^{(1)}(k_1,k_2)=i \}|,
\end{eqnarray}
where $\pi_{b_k}^{(1)}(k_1,k_2)$ denotes the first coordinate of
$\pi_{b_k}(k_1,k_2)$. Further, we define for each $1 \leq i \leq r$ a
subset $B_i \subset \{0,1\}^{m}$ by
\begin{eqnarray}
	B_i:=\{\bb:\ n_i(\bb)&>& n_j(\bb)\textrm{ for } 1\leq
	j<i,\nonumber \\ n_i(\bb) &\geq& n_j(\bb)\textrm{ for } i\leq
	j\leq r\}.
\end{eqnarray}
Finally, we set
\begin{eqnarray}\label{def:E_i}
	E^{(n)}_i:=\sum_{\bb \in B_i} P^{(n)}_{\bb}.
\end{eqnarray}  
Note that $B_i \cap B_j= \emptyset$, for $i\not= j$, and $\bigcup_{i=1}^r B_i=
\{0,1\}^{m}$, i.e. $\{B_i\}_{i=1}^r$ represents
a (disjoint) decomposition of the set $\{0,1\}^{m}$ of binary
sequences of length $m$. Hence $\{E^{(n)}_i\}_{i=1}^r$ defines a POVM
in $\mathcal{A}^{(n)}$, and associating the measurement outcome
corresponding to $E_i^{(n)}$, $i=1,\dots, r$, to the density operator
$\rho_i^{(n)}$, respectively, we obtain a proper quantum test for
$\Sigma^{(n)}$.
\\
\\
It remains to verify the asymptotic behaviour (\ref{bound-thm}) for
$E^{(n)}$. To this end, we fix an $i \in \{1,\dots, r\}$, define a
corresponding index set
\[
	K_i:=\{k\in \{1,\dots, m\}:\ k_1=i \textrm{ or }
	k_2=i\},
\] 
and consider the individual error probability
$\textrm{Err}_i(E^{(n)})$. We have
\begin{eqnarray}
	\textrm{Err}_i(E^{(n)})&=& \textrm{tr }[\rho_i^{(n)} (\idn_n -
	E_i^{(n)})]\nonumber \\ &=&\textrm{tr }[\rho_i^{(n)} \sum_{j\not=
	i}E_j^{(n)}]\nonumber \\ &=&\textrm{tr }[\rho_i^{(n)}
	\sum_{\bb \notin B_i}P_{\bb}^{(n)}]\nonumber \\ &\leq& \sum_{k
	\in K_i}\textrm{tr }[\rho_i^{(n_k)}
	P_{i^{\perp}_k,i}^{(n_k)}]=\sum_{1\leq j\leq r, j\not=
	i}\textrm{tr }[\rho_i^{(n(i,j))}
	P_{j,i}^{(n(i,j))}],\label{inq}
\end{eqnarray}
where the first line is by definition of individual error probability
and the second line is according to (\ref{def:E_i}). The index
$i_k^{\perp}$ appearing on the right hand side of inequality
(\ref{inq}) is such that the subblock $\I_k$ corresponds to the pair
of density operators $\rho_i^{(n_k)}$ and
$\rho_{i_k^{\perp}}^{(n_k)}$. Inequality (\ref{inq}) follows from the
fact that at least one tensor factor of a projector $P_{\bb}^{(n)}$
with $\bb\notin B_i$ is equal to a Holevo-Helstrom projector of the
form $P_{j,i}^{(n(i,j))}$, where $j\not= i$, i.e. $P_{j,i}^{(n(i,j))}$
corresponds to decision in favor of $\rho_j^{(n(i,j))}$ and against
$\rho_i^{(n(i,j))}$. More in detail, we deduce the inequality as
follows. For any $\bb \notin B_i$ there exists an index $k \in K_i$
with $\pi_{b_k}^{(1)}(k_1,k_2)\not= i$. Let $k_{\bb}$ be the smallest
of such indices corresponding to $\bb$. We denote by $B_i^{\perp}(k)$
the set consisting of all $\bb\notin B_i$ with $k_{\bb}=k$:
\[
	B_i^{\perp}(k):=\{\bb \notin B_i:\ k_{\bb}=k\}.
\]
Observe that $\{B_i^{\perp}(k)\}_{k \in K_i}$ represents a
decomposition of $B_i^{\perp}:=\{0,1\}^{m}\setminus B_i$ into $r-1$
disjoint subsets. For each $k\in K_i$ we deduce the following upper
bound on the sum of projectors $P_{\bb}^{(n)}\in \mathcal{A}^{(n)}$
over $B_i^{\perp}(k)$ in terms of the projector
$P_{i^{\perp}_k,i}^{(n_k)}$, which is understood here as an element in
the local algebra $\mathcal{A}_{\I_k}\subseteq
\mathcal{A}_{\I}=\mathcal{A}^{(n)}$:
\begin{eqnarray}
	\sum_{\bb \in B_i^{\perp}(k)} P_{\bb}^{(n)} \leq
	P_{i^{\perp}_k,i}^{(n_k)}\otimes \idn_{\I \setminus \I_k}.
\end{eqnarray}
The index $i^{\perp}_k$ is again determined by $k$ as explained below
(\ref{inq}), and $\idn_{\I \setminus \I_k}$ denotes the identity in
the local algebra $\mathcal{A}_{\I \setminus \I_k}\subset
\mathcal{A}^{(n)}$ associated to the subset $\I
\setminus \I_k$ of $\I$. It follows the estimate
\begin{eqnarray}
	\sum_{\bb \notin B_i}P_{\bb}^{(n)} =\sum_{k \in K_i}\sum_{\bb
	\in B_i^{\perp}(k)} P_{\bb}^{(n)} &\leq& \sum_{k \in K_i}
	P_{i^{\perp}_k,i}^{(n_k)}\otimes \idn_{\I \setminus
	\I_k}\nonumber,
\end{eqnarray}
which, applying the shift-invariance of $\omega_i$, implies the upper
bound (\ref{inq}) on $\textrm{Err}_i(E^{(n)})$.
\\
\\
Assume now that all subblock lengths $n(i,j)$, $i\not= j$, are
(asymptotically) proportional to $n$ with factor $w_{ij}$, i.e.
\begin{eqnarray}
n(i,j) &  =&w_{ij}\;n\;\left(  1+o(1)\right),\nonumber\\
\sum_{1\leq j<i,1\leq i\leq r}w_{ij} &  =&1.\label{restric}
\end{eqnarray}
Recall that for each pair $(i,j)$ of indices the $P_{j,i}^{n(j,i)}$
in (\ref{inq}) denote the Holevo-Helstrom projectors corresponding to
the two density operators $\rho_{i}^{(n(i,j))}$ and
$\rho_{j}^{(n(i,j))}$, and hence they represent a sequence of
(asymptotically) optimal quantum tests for $\{\omega_i,\omega_j\}$
achieving the asymptotic error exponent equal to the mean quantum
Chernoff distance $\bar \xi_{QCB}( \omega_i, \omega_j)$.  Hence we
obtain from (\ref{inq}) as $n$ tends to infinity
\begin{eqnarray}
	-\liminf_{n\rightarrow\infty}\frac{1}{n}\log\mathrm{Err}_{i}(E^{(n)})\geq
	\min_{i\neq j}w_{ij}\bar
	\xi_{QCB}(\omega_i,\omega_j).\label{estim-2}
\end{eqnarray}
Note that the minimum on the right hand side appears due to the fact
that asymptotically the largest term in (\ref{inq}) dominates.

In order to get the best lower bound, i.e. to maximize the right hand
side of (\ref{estim-2}) under the restriction (\ref{restric}), we
solve the problem
\[
	\max_{w_{ij}}\left\{ \min_{i\neq j}w_{ij}\bar
	\xi_{QCB}(\omega_i, \omega_j):\sum_{1\leq j<i,1\leq i\leq
	r}w_{ij}=1\right\}.
\]
The solution is obtained by making all $w_{ij}\bar \xi_{QCB}(\omega_i,
\omega_j)$ equal, that is by setting
\[
	w_{ij}=\frac{1}{\left( \bar \xi_{QCB}(\omega_i,
	\omega_j)\sum_{1\leq t<s,1\leq s\leq
	r}\frac{1}{\bar\xi_{QCB}(\omega_s,\omega_t) }\right) }
\]
whence
\[
	\min_{i\neq
	j}w_{ij}\bar\xi_{QCB}(\omega_i,\omega_j)=\frac{1}{\sum_{1\leq
	j<i,1\leq i\leq r}\frac {1}{\bar
	\xi_{QCB}(\omega_i,\omega_j)}}.
\]
The mean generalized quantum Chernoff bound of the set $\Sigma$ was defined
as $\bar\xi_{QCB}(\Sigma)=\min_{i\neq
j}\bar\xi_{QCB}(\omega_i,\omega_j)$, and we obtain
\begin{eqnarray}
	-\liminf_{n\rightarrow\infty}\frac{1}{n}\log\mathrm{Err}_{i}(E^{(n)})\geq
	\frac{1}{\sum_{1\leq j<i,1\leq i\leq
	r}\frac{1}{\bar\xi_{QCB}(\omega_i,\omega_j)}}=\bar
	\xi_{QCB}\cdot
\phi(\Sigma),  \label{improved-bound}
\end{eqnarray}
where the factor on the right hand side is given by
(\ref{def:phi-factor}), i.e.  $\phi(\Sigma) =\left(\sum_{1\leq
j<i,1\leq i\leq
r}\frac{\bar\xi_{QCB}}{\bar\xi_{QCB}(\omega_i,\omega_j)}\right)^{-1}$.
Since the lower bound (\ref{improved-bound}) on the individual error
exponent does not depend on the index $i$, $i=1,\dots, r$, the
statement of the Theorem \ref{thm:exp-decay}, which refers to the
exponential rate of the averaged error probability, follows. \qed
\\
\\
\textbf{Acknowledgements.} 
A.~S. likes to thank the members of the groups of Nihat Ay
and J\"urgen Jost at the MPI MiS, Mil\'an Mosonyi and Markus M\"uller
for their interest in the topic and useful discussions. The work of
M.~N. was supported in part by NSF Grant DMS-03-06497.

\end{document}